\pdfoutput=1

\documentclass[11pt]{article}

\usepackage{acl}

\usepackage{times}
\usepackage{latexsym}

\usepackage{placeins}
\usepackage[T1]{fontenc}

\usepackage[utf8]{inputenc}

\usepackage{microtype}

\usepackage{hyperref}
\usepackage{url}
\usepackage{booktabs} 
\usepackage{caption,setspace}
\usepackage{balance}
\usepackage{graphicx}
\usepackage{adjustbox}
\usepackage{xcolor}
\usepackage{multirow}
\usepackage{float}
\usepackage{algpseudocode}
\usepackage{algorithm}
\usepackage{subcaption}

\usepackage{amsmath}

\title{Intersectionality and Testimonial Injustice in Medical Records}

\author{Kenya Andrews, Bhuvani Shah, Lu Cheng\\
  kandre32@uic.edu | bshah46@uic.edu | lucheng@uic.edu\\
  University of Illinois at Chicago \\ Computer Science}

\begin{document}
\maketitle
\begin{abstract}
Detecting testimonial injustice is an essential element of addressing inequities and promoting inclusive healthcare practices, many of which are life-critical. However, using a single demographic factor to detect testimonial injustice does not fully encompass the nuanced identities that contribute to a patient's experience. Further, some injustices may only be evident when examining the nuances that arise through the lens of intersectionality. Ignoring such injustices can result in poor quality of care or life-endangering events. Thus, considering intersectionality could result in more accurate classifications and just decisions. To illustrate this, we use real-world medical data to determine whether medical records exhibit words that could lead to testimonial injustice, employ fairness metrics (e.g. demographic parity, differential intersectional fairness, and subgroup fairness) to assess the severity to which subgroups are experiencing testimonial injustice, and analyze how the intersectionality of demographic features (e.g. gender and race) make a difference in uncovering testimonial injustice. From our analysis, we found that with intersectionality we can better see disparities in how subgroups are treated and there are differences in how someone is treated based on the intersection of their demographic attributes. This has not been previously studied in clinical records, nor has it been proven through empirical study. 
\end{abstract}

\section{Introduction}
\label{sec:intro}
In medical settings, decisions can have life-critical consequences \citep{ZENIOS199952, KUMARMANGLA2023113398, ventcritical,cheng2021socially,cheng2023socially}, making it \emph{essential} to ensure that machine learning tools use there are fair. This fairness is often measured with common fairness metrics such as demographic parity \citep{dwork2012fairness} and equal opportunity \citep{hardt2016equality}. However, these tools do not consider the intersectionality of the subjects under consideration \citep{intersectional-comparisons,Gohar2023intersection}. That is, by focusing solely on factors such as race, gender, or socioeconomic status, we ignore the nuances related to individuals with unique experiences shaped by having multiple features sensitive to marginalization. We theorize that \emph{how various aspects of an individual intersect and contribute to their experiences, via intersectionality, could make instances of injustice more overt - and in some cases may be the sole approach for identifying such instances}. Intersectionality recognizes that power relations based on factors such as race, class, and gender are not mutually exclusive and can interact with each other, affecting all aspects of the social world \citep{marques2018patricia}. Therefore, it is important to consider intersectionality when evaluating the fairness of machine learning tools in medical settings.

In clinical settings, it is particularly important that care providers (e.g. physicians) properly acknowledge what their patients are hoping to convey to them in a way that does not diminish what the patient is saying. Moreover, it is imperative for care providers to accurately relay their understanding of their patients' experiences, as others will be dependent upon their previous understandings and evaluations, often recorded in notes, to assist with overseeing and providing care for that patient \citep{notes}. We have seen that when this does not occur, there are higher instances of death amongst certain marginalized groups \citep{bowman2013impact}. With the rise in using machine learning tools to help make decisions on medical plans and treatments, who often only interact with the notes provided to them and not the actual patient, it is vital they are able to properly see patients. This visibility should be clear despite previous attempts at burying their words behind instances of injustices which hides them as a speaker. Here, we focus on a particular form of injustice - testimonial injustice. Testimonial injustice occurs when someone is assigned less credibility due to prejudices about them \citep{fricker2019testimonial}. 

The aim of our study is to examine how testimonial injustice in medical records is affected by the intersectionality of gender and race. These two observable attributes have historically led to marginalization in various societal settings, such as education \citep{educationracegen}, housing \citep{roscigno2009complexities}, and healthcare \citep{KRIEGER19901273, implicit-bias}. In fact, some forms of marginalization may only be evident in those with multiple marginalized identities - for instance, a Black police woman may not experience the same level of power and privilege as a White male police officer \citep{martin1994outsider}. Neglecting to consider the various contributing identities of an individual may further marginalize them. Therefore, it is important to consider intersectionality when identifying and addressing injustices in order to result in more accurate classifications and decisions. 

There has been a small amount of work done to understand testimonial injustice in medical records and to our knowledge no prior work on how intersectionality might affect the emergence of testimonial injustice, even in life-critical medical settings. This motivates our contributions to this work: (1) The importance of intersectionality has been spoken about but has not been shown before (particularly in the medical setting). Thus, we perform an empirical study to show there is a difference in how subgroups are treated in medical settings, but this can only be revealed in intersectional views. (2) Practitioners continue to use singular-feature fairness metrics in medical settings. Thus, we provide proof that we should not be using these metrics to detect instances of injustice. This proof has not been provided before, not even in medical settings. Thus, we (3) perform an empirical study to show traditional fairness metrics (i.e. demographic parity) are inefficient when judging people's experiences in healthcare because they produce different results when the entirety of a person is considered. (4) Lastly, not all metrics fit each situation - even in similar settings. Therefore, we analyze if different intersectional fairness metrics might reveal differences in how we recognize intersectionality.

Previous studies have shown that both Black patients and female patients are more likely to experience testimonial injustice in the medical field, as evidenced by the use of biased language in their records \citep{beach2021testimonial}. However, these studies have not examined the specific impact of intersectionality, or how being simultaneously Black and female might affect testimonial injustice. Our work seeks to address this gap by examining the impact of the intersection of ethnicity (Black, Asian, Latino, and White) and gender (Male and Female - though we acknowledge in modern society, there is recognition of genders beyond the traditional binary options, the dataset used here only includes these two genders) on testimonial injustice in medical records.



\section{Related Works}
Despite the increased use of machine learning tools and a growing focus on intersectionality in the medical community \citep{inter-perspectives, pubhealth}, there have been limited efforts to understand how intersectionality can impact outcomes in medical settings. Since various healthcare professionals rely on medical records to make treatment decisions and give proper care, it is crucial that such records are written appropriately \citep{bali2011management}. The authors of \citep{ham22} found that even when race is removed from patients' records, models could detect the race of the patient - even when humans could not. Furthermore, they discovered that models trained on these records (i.e. which race has been removed from) still maintain biases in treatment recommendations. Though they only remove race in their work, this further affirms that there are differences in how patients are spoken about in their records based on demographic features, emphasizing the need to study what can occur if we look at multiple demographic features as we do here. In their work, \citeauthor{stig-effects} explored how stigmatizing language in a patient's medical record can shape the attitudes of physicians-in-training towards the patient and their clinical decision-making. They found that stigmatizing language is associated with more negative attitudes and less aggressive pain management. Building on this work, we examine words that may indicate testimonial injustice, which occurs when someone's statements are diminished due to stereotypes or prejudices about them \citep{fricker2019testimonial}. It is therefore important to identify instances of stigmatizing language in medical records and take steps to prevent them from occurring as emphasized by \citeauthor{park21}. 

In \citep{beach2021testimonial}, the authors use a lexicon look-up to identify testimonial injustice in medical records, analyzing the use of quotation marks, evidential words, and judgmental words in the records of male and female patients who are Black or White. We expand their work, including words that are negative and commonly used stigmatizing words in medical settings. We exclude the search for quotation marks, acknowledging that direct quotations may give rise to uncertainty by suggesting that the statement in question constitutes not a fact, but rather an assertion \citep{beach2021testimonial}. However, we believe that our expanded lexicon will help to identify instances of testimonial injustice. Further in contrast to \citeauthor{beach2021testimonial}, we consider the records of Black, White, Asian, and Latino patients, exploring how testimonial injustice may differ across the intersection of their identities with gender. The authors found that Black and female patients are most likely to experience testimonial injustice, highlighting the need to examine how different intersectional identities impact experiences of testimonial injustice in medical settings.

Previous research has examined the presence of epistemological bias in medical records based on sensitive attributes to detect instances of experiences injustice i.e. disparate treatment. \citeauthor{stigmatizing} studied diabetic patients and found that non-Hispanic Black patients were more likely to have stigmatizing language included in their notes than non-Hispanic White patients. Similarly, \citeauthor{neg-words} investigated medical records and racial bias, discovering that Black patients had a 2.54 times higher chance of negative descriptors than White patients. These studies suggest that certain demographics may experience differential treatment in medical settings, which may help explain healthcare disparities. However, these works only examined single demographic features, while we seek to investigate their intersection. We anticipate that studying the intersection of groups will more clearly reveal instances of injustice or discrepancies in treatment. The ongoing use of tools that do not consider intersectionality highlights the importance of this research \citep{gender-shades}. 

\citeauthor{word-embeddings} developed a technique to automatically identify intersectional biases from static word embeddings. They found that their model's highest accuracy was for predicting emergent intersectional bias among African American and Mexican American women. This could be attributed to these groups experiencing more overt biases that are easier to detect. This discovery motivates us to further investigate if biases are more prevalent in high-risk settings such as medical settings, especially for individuals from marginalized groups. However, it can be challenging for humans to identify when a bias is occurring since it can be subtle, as highlighted by \citeauthor{wikibias}. Furthermore, doctors may struggle to recognize their own use of words that cause testimonial injustice since they may be unconsciously influenced by their own biases and take them as facts \citep{fitzgerald2017implicit,beeghly2020introduction}.
\section{Data}
\subsection{MIMIC-III}
Obtaining medical data has been a standing challenge, largely due to HIPAA requirements and privacy constraints. We use the MIMIC-III \citep{mimciii} dataset, which contains features of interest to our experiments: ethnicity/race, gender, patient id, diagnosis, physicians' notes , and so on. This data was collected between 2001-2012 at the Beth Israel Deaconess Medical Center in Boston, MA. The MIMIC-III dataset contains information for 46,146 patients. The distribution of racial groups in the data was highly disproportionate, as shown in Table \ref{tab:distribution}. The two genders represented in this dataset, Female and Male, however, are more balanced. We removed ethnicities that were listed as ``unknown/not specified", ``multi-race ethnicity", ``other", ``unable to obtain", and ``patient declined to answer" since we cannot clearly denote the race of these patients. We also removed patients whose diagnosis was ``newborn" since these patients had notes solely stating they were newly born. We did however include the newborns who had other diagnoses. Only 9 of those patients were Caribbean and 38 were Middle Eastern, thus we removed them from the records as well. We were not able to find any duplicate records in the dataset, with a simple python search.

After data pre-processing, there are 32,864 patients in total for experimentation. We truncated the MIMIC-III feature 'ethnicity' into 'race' such that all ethnicities are represented as the race often associated with them as labeled in the dataset (e.g. original ethnicity in the dataset: 'ASIAN - VIETNAMESE' was truncated to 'Asian'). For ethnicities that were not associated with a particular race, we searched for how they are commonly associated and relabeled them to the race (e.g. original ethnicity in the dataset: 'SOUTH AMERICAN' was relabeled to 'Latino'). Finally, given that many patients had multiple records, we clustered the patients based on their patient\_id and combined their records based on patient\_id, gender, race, and diagnosis (e.g. 56327, male, Latino, HYPOTENSION). We then run analysis on the physicians' notes to find terms that are testimonially injust.

\begin{table}[h]
\fontsize{11pt}{12pt}\selectfont
\centering
\begin{tabular}{l|l|l}
\hline
\textbf{Race}   & \textbf{Gender} & \textbf{Count} \\ \hline
White  & Female & 15,399 \\ \hline
Black  & Female & 2,522  \\ \hline
Asian  & Female & 512   \\ \hline
Latina & Female & 662   \\ \hline
White  & Male   & 20,317 \\ \hline
Black  & Male   & 2,041  \\ \hline
Asian  & Male   & 690   \\ \hline
Latino & Male   & 1,041  
\end{tabular}
\caption{Counts of patients by race and gender. \label{tab:distribution}}
\end{table}

We analyze the distribution of data for MIMIC-III in \ref{ssec: dist-analysis}. Our analysis looked at the occurrence of our four types of words associated with testimonial injustice, namely evidential (Figure \ref{fig:evi-graphs-append} and \ref{fig:evi-graphs}), judgmental words (Figure \ref{fig:judge-graphs-append} and \ref{fig:judge-graphs}), stigmatizing words (Figure \ref{fig:stig-graphs-append} and \ref{fig:stig-graphs}), and negative words. We plot the density distribution of each gender, race, and their intersection as normalized sums of these types of words, where the numerator is the frequency of occurrence of the relevant words for that patient and the denominator is the number of records for that patient.  We did not include the plots for negative words due to their limited occurrence in the medical notes of this dataset, however we do use them in our analysis of the results for detecting testimonial injustice. Our observations suggests that the confluence of race and gender better helps us in distinguishing instances of testimonial injustice than either race or gender in isolation. In particular, when race and gender are considered independently, males seem to be treated better than females or White patients are treated generally better than Black patients. However, there is nuance in the difference in the treatment of White males and White females as well as Black males and Black females. 

\subsection{Testimonial Injustice Terms}\label{ssec: terms}
In order to assess testimonial injustice in the physicians' notes, we focus on 4 main categories of unjust words: evidential, judgemental, negative, and stigmatizing words that can contribute to someone experiencing testimonial injustice. 
We use the same evidential and judgmental words from \citep{beach2021testimonial}. Evidential terms do not endorse a statement but allow it to be agnostic (e.g. ``complains", ``says", ``tells me" and so on). When a physician uses these words, they express dismissing what the patient is actually experiencing. Judgment terms cast doubt on the sayer by the hearer (i.e. the physician) by trying to make their statements sound good or bad (e.g. ``apparently", ``claims", ``insists", and so on). Exacerbated racial and ethnic healthcare disparities have been linked to negative words used to describe Black patients as well \citep{neg-words}. Negative words are included in this study as they typically show active rejection or disagreement, e.g. ``challenging", ``combative", ``defensive", ``exaggerate", and so on. Clearly, the use of these words expresses assumptions about the patient and could result in a lower quality of care. 

We also include stigmatizing terms as they are commonly used in medical contexts \citep{stigmatizing}. Stigmatizing terms are rooted in stereotypes or stigmas about a person \citep{link2001conceptualizing} (e.g. ``user", ``faking", ``cheat", and so on). Using stigmatizing terms may alter treatment plans, transmit biases between clinicians, and alienate patients. This lexicon has been proven to consist of words used to diminish specific conditions like diabetes, substance use disorder, and chronic pain \citep{stigmatizing}. All of these conditions are known to disproportionately affect racial minority groups. Using all of these terms in our lexicon lookup \ref{ssec: lexicon-lookup} will help us to detect testimonial injustice in these medical records.

\section{Methods}
Although all marginalized groups invariably experience some degree of injustice, our aim is to bridge the gap in research by highlighting the disparate treatment of subgroups in medical notes.
To achieve this goal, we estimate and compare common metrics across different groups (i.e. Asian men, Asian women, Black men, Black women, Latino men, Latina women, White women, and White men) specifically using demographic parity, differential intersectional fairness, and subgroup fairness.

\subsection{Normalization}
To account for patients who had multiple visits or were admitted to the ICU for multiple days, the physicians' notes were combined for each patient's duration in the ICU. To analyze the potential variance in testimonial injustice among different groups, we summed the frequency of testimonial injustice words in the notes for each patient and then normalized this frequency by dividing it by the number of original records we had for that particular patient. This allowed us to ensure that each patient had an equal standing, regardless of length of hospital stay or number of visits from doctors. By using normalized sums, we were able to compare groups and determine if there were any differences in levels of testimonial injustice. The normalized sums of occurrences of testimonial injustice across each intersection of groups are visualized in Figure \ref{fig:interscores} in \ref{ssec: normsumappendixsec}.

\subsection{Lexicon Lookup} \label{ssec: lexicon-lookup} 
After normalizing the sums of testimonial injustice for each patient, we performed a lexicon lookup for exact phrase matching. With this, we counted the frequency of occurrence for each testimonial injustice word in the patients’ combined and normalized visits. We combined the terms introduced in Section \ref{ssec: terms} commonly associated with being evidentially biased, judgmental, negative, and stigmatizing into a lexicon. 

\subsection{Defining Fairness} \label{ss: def fairness}
In this work, we define the desired fairness as the following: \textit{a patient's record has \textbf{no} terms which are considered testimonial unjust.} However, this is a strict boundary that is unlikely to be met since a term could appear in a patient's record but might not actually be casting doubt on them as a sayer (i.e. testimonial injustice). Thus, we find the greatest number of occurrences of each type of term that indicates testimonial injustice, $m = max_p(t/r)$ (where $p$ are the patients). We determine that if a patient has more than $m*.10$ in that particular type of term, they as experiencing testimonial injustice. For this work, we arbitrarily use 10\% of the maximum value for each term. In the future, we will do some experimentation to improve this definition of fairness. To determine if there is disparate treatment amongst groups to this fairness definition, we use fairness metrics - demographic parity, differential intersectional fairness, and subgroup fairness.

\subsubsection{Demographic Parity}
Demographic parity requires that the difference in two groups being assessed have equal chances of receiving a positive outcome \citep{dwork2012fairness}. We use this metric as our baseline metric to understand how testimonial injustice might reveal itself if we ignore intersectionality, as has been done with most works in the fairness literature [\citep{hardt2016equality}, \citep{kusner2017counterfactual}, \citep{agarwal2018reductions},and so on]. That is, we are seeking to investigate whether there is a significant difference in the way a patient is spoken about in medical records when the intersection of their race and gender are considered. Demographic parity is a popular fairness metric, but it does not work to reveal fairness or justice; rather it solely reveals equity. We can look at the example of when both groups have high amounts of injustice (i.e. true fairness occurs when neither group experiences injustice, nearly 0) hence, fairness is not detected only equality or when a marginalized group should be afforded more opportunity for the sake of corrective justice due to historical bias hence justice is not enforced. In these cases, demographic parity is still satisfied, but fairness nor justice persists. Demographic parity is defined as:

\begin{equation}
    \frac{P(Y=1|A=a)}{P(Y=1|A=a')} > 0.8,
\end{equation}
\noindent where $Y$ is the outcome and $A$ is the sensitive attribute. Demographic parity looks to ensure the difference between the two groups receiving a positive outcome is greater than 80\%.

\subsubsection{Differential Fairness}
For intersectionality, we first look at $\epsilon$-Differential fairness \citep{intersectional-fairness}, which requires that the difference between groups, regardless of their combination of sensitive attributes, not be treated differently within a range. This metric of fairness allows us to include multiple attributes of a person whereas demographic parity only allows us to look at one sensitive attribute per group. Differential fairness is defined as:
\begin{equation}
e^{-\epsilon} < \frac{P(M(x)=y|s_i,\theta)}{P(M(x)=y|s_j,\theta)} < e^{\epsilon},
\end{equation}
where $\epsilon$ should be small. In our experiments, it is set to 0.01, $M$ is a mechanism (linear regression in our case) that takes an instance, $x$, from the data to achieve some outcome, $y$, $s$ values are the cross product of sensitive attributes, and $\theta$ is the distribution of $x$.

\subsubsection{Subgroup Fairness}
Another common intersectional fairness notion is Statistical Parity Subgroup Fairness or subgroup fairness. We use subgroup fairness to compare our results with the differential fairness metric. Subgroup fairness \citep{subgroup} requires there be no difference in positive outcomes between groups, but we are allowed to ignore an $\alpha$ amount of people. Subgroup fairness is described for each group, $a$, by:
\begin{equation}
    \alpha(a,\mathcal{P})*\beta(a, M, \mathcal{P}) \leq \gamma ,
\end{equation}
where,
\begin{equation*}
    \begin{split}
        \alpha(a,\mathcal{P}) = P_\mathcal{P}[a(x)=1]\\
        \beta(a, M, \mathcal{P}) = |P_{\mathcal{D}, \mathcal{P}}[M(x)=1] - \\P_{M, \mathcal{P}}[M(x)=1 | a(x) = 1]|.
    \end{split}
\end{equation*}

\noindent Here $M$ is a classifier, $\mathcal{P}$ is the distribution of patients, $\gamma \epsilon [0,1]$ indicates the amount of deviation from equity we tolerate. We relax this constraint for our experiments, allowing $\gamma$ to be 95\% of the maximum value of $\alpha(a,\mathcal{P})*\beta(a, M, \mathcal{P})$ for each term that leads to testimonial injustice. $a(x) = 1$ indicates that individuals with sensitive feature, $x$, are in group $a$.

\section{Results} \label{sec:results}
When examining the results for demographic parity, we solely focus on instances of race or gender, as this approach only allows for an assessment of one factor at a time. However, for differential fairness and subgroup fairness, we conduct an intersectional analysis with race and gender. For these, we look to see which groups have privilege over another, meaning one group experiences less testimonial injustice in their physicians' notes as opposed to the group they are being compared to.
\subsection{Demographic Parity}
\begin{figure}[h]
\includegraphics[width=7.5cm]{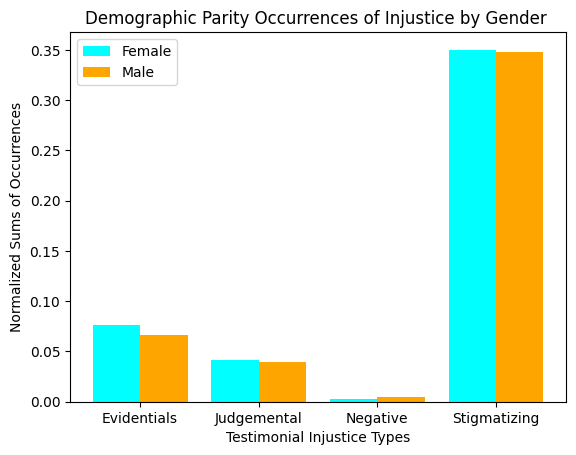}
    \caption{Demographic Parity Occurrences of Injustice by Gender.}
    \label{fig:occur_inj_demo_gen}
\end{figure}

\noindent\textbf{Gender.} In terms of Demographic Parity gender analysis, there was little to no disparate treatment detected across all term types between male and female patients, indicating that there was minimal evidence of injustice in the data based on gender, as observed in Figure \ref{fig:occur_inj_demo_gen}. The greatest difference was found within evidential words, where female patients experienced the most injustice. Then follows the stigmatizing words and judgment words with the greatest bias against females. The least difference comes from the negative words with males experiencing the least fairness. Negative words occurred the least and stigmatizing words occurred the most across the patient records. With this, gender should not be found to be a significant predictor of the treatment or care received by patients. Therefore, the findings of the analysis should show that a person’s gender membership does not have any substantial impact on how they are treated, indicating that the principle of fairness is being upheld.

\begin{figure}
\includegraphics[width=7.9cm]{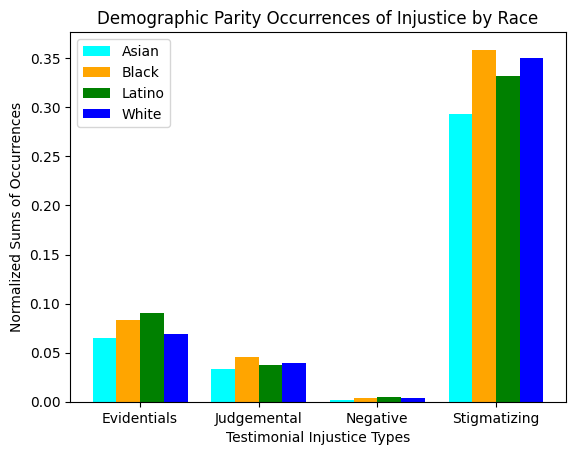}
    \caption{Demographic Parity Occurrences of Injustice by Race.}
    \label{fig:occur_inj_demo_race}
\end{figure}

\noindent\textbf{Race.} In terms of Demographic Parity race analysis, there was little to no disparate treatment detected across all term types between  the different races of patients, indicating that there was minimal evidence of injustice in the data based on race, as observed in Figure \ref{fig:occur_inj_demo_race}. We observe that Latino patients are the most likely to experience evidential words, while Asian patients were the least likely. Further, for evidential words, White patients have privilege over Black patients, Black patients have privilege over Latino patients, and Asian patients have privilege over White and Latino patients. For judgemental words, Black patients are the most likely, and Asian patients were the least likely to experience judgemental words. Here, we observe that White patients have privilege over Black patients. Latino patients were the most likely and Asian patients were the least likely to experience negative words in their medical records. We note here that negative terms were the least likely to appear in the records of any patient. Black patients were the most likely and Asian patients were the least likely to experience stigmatizing words in their medical records. Another observation is that White patients have privilege over Black patients, Asian patients have privilege over every race of patients, and Latino patients have privilege over White patients. Stigmatizing words occurred the most in everyone's medical records. With this, race should also not be found to be a significant predictor of the treatment or care received by patients. Therefore, the findings of the analysis should show that a person’s racial membership does not have any substantial impact on how they are treated, indicating that the principle of fairness is being upheld.

Since our analysis using demographic parity showed that neither race nor gender affect how a patient experiences testimonial injustice, when we observe their intersection, we should see that the treatment and care received by patients are not affected by the intersectionality of race and gender. This would indicate that the principle of fairness is being upheld regardless of a patient's race or gender. However, we see a different story when we consider intersectionality.

\subsection{Differential Fairness}
Differential fairness focuses on the intersectionality of race and gender in relation to testimonial injustice. The results of the demographic parity experiments showed, there are no disparities in how groups are treated with respect to testimonial injustice upon race or gender. However, the results of the experiment pertaining to differential fairness show that there are disparities between different intersections of gender and race with respect to the types of terms that lead to testimonial injustice. Specifically, out of 112 comparisons for each intersection of gender and race, 110 violations of differential fairness occurred. This demonstrates that there are underlying injustices occurring in how different groups are treated based on gender and race and that we cannot simply rely on measures that do not consider intersectionality to reveal this.

\begin{figure}[h]
\includegraphics[width=7.5cm]{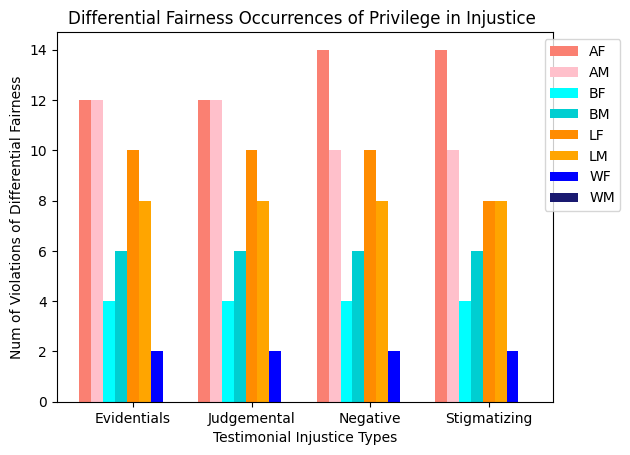}
    \caption{Differential Fairness Occurrences of Injustice by Gender and Race.}
    \label{fig:occur_inj_diff_int_winners}
\end{figure}

There were very few instances in which fairness was not violated, such as Asian males to Asian females for evidential and judgmental words, and Asian males to Latina females for negative words. The results showed that Asian females and males were the most privileged, and White males and females were the least privileged when fairness was violated. This may be due to the fact that there are many more records for White patients than all other races of patients. As observed in Figure \ref{fig:occur_inj_diff_int_winners}, across all types of terms that lead to testimonial injustice, Black females were the next least privileged after White patients. Black males were found to have more privilege in experiencing testimonial injustice than Black females. The experiment was also conducted with 500 randomly sampled records of each subgroup of patient, and the results there showed that when unfairness is present, Black females are the most marginalized, and Asian males are the least. For these sampled records, across all types of terms that lead to testimonial injustice, Latina females were the most marginalized for evidential words, Black females for judgment words and negative words, and Latino males for stigmatizing words. However, even with the full dataset, Asian males were consistently found to be the most privileged of all the groups represented. 

\subsection{Subgroup Fairness}
In this experiment, similar to differential fairness, we focus on the intersectionality of race and gender in relation to testimonial injustice. The results of the demographic parity experiments showed, there were no disparities in how groups were treated with respect to testimonial injustice upon race nor gender. However, the results of the differential fairness experiments showed there are differences in how one is treated based on their race and gender. We conduct an experiment that also looks at intersectionality of groups to compare if there is a difference in how these two metrics reveal disparate treatment amongst the subgroups.

\begin{figure}[h]
\includegraphics[width=7.5cm]{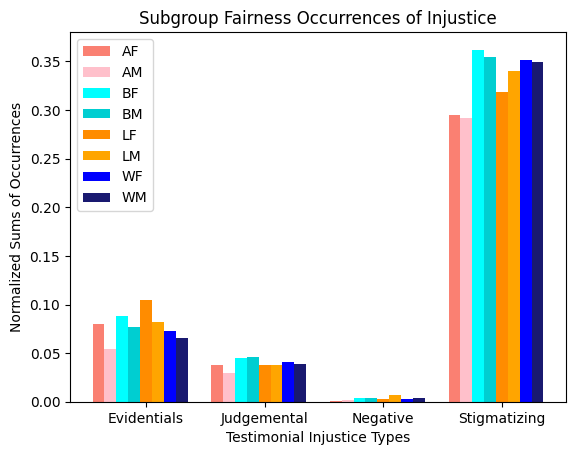}
    \caption{Subgroup Fairness Occurrences of Injustice by Gender and Race.}
    \label{fig:occur_inj_sub_int}
\end{figure}

Based on our analysis of demographic parity in detecting testimonial injustice in medical records, we found that the privileged groups by race are Asian and White patients, as well as males. Therefore, for the purpose of intersectional fairness analysis, we consider Asian men and White men as non-sensitive groups. When we conducted a differential fairness analysis, we found that violations occurred 110 times out of 112 comparisons (each intersection of gender and race for each type of term leading to testimonial injustice). We expected similar results (Figure \ref{fig:occur_inj_sub_int}) for subgroup fairness analysis. Our subgroup fairness metric detected 69 violations our of the 112 comparisons of subgroups. Though less occurrences of violations are present, this still reveals we must consider intersectionality within the medical setting and in the fairness metrics we use there. If even better highlights that a metric which considers intersectionality is not enough, but we must be careful at which fairness metrics we use based on the tasks at hand.

For evidential terms, we found that Latina females were the most discriminated against, while Asian males were the most privileged. For judgment terms, Black males were the most discriminated against, while Asian males were the most privileged. For negative words, Asian males were the most privileged, while Latino males were the least privileged. For stigmatizing words, Black females were the most discriminated against, while Asian males were again the most privileged. It is important to note that our experiment includes the entire dataset, which is over-representative of White patients. Thus, we can expect even larger disparities in how different groups are treated with a more representative dataset. This does not mean that White patients do not experience discrimination, but rather emphasizes the importance of having a more representative dataset to better understand the degrees to which different groups may experience testimonial injustice in their records.

\section{Discussion}
When conducting experiments using demographic parity, we compared race or gender. In each case, there were no violations of demographic parity for any patient is treated based on their race or gender alone. If a practitioner takes these results for face value, they might determine there is no form of discrimination happening based on these commonly observed visible attributes. For example, when speaking to a Black male patient who was stigmatized against from the demographic parity view, they would have no evidence in that setting to back their expression of their experience. However, when we look deeper, through the lens of intersectional fairness (i.e., differential fairness and subgroup fairness) at the intersection of race and gender, we can see that a male patient can still experience discrimination (i.e. Black males) and so could a White patient (i.e. White females). 

When we look at measures that consider intersectionality, we see disparity in how people are treated based on their race and gender for every type of word we analyzed that could lead to testimonial injustice. We attribute this to: (1) being able to consider multiple aspects about a person that might only reveal themselves at the intersection of race and gender, (2) in differential fairness being able to constrain the range in which we look for violations, as opposed to only looking at it from one side as demographic parity does. To properly see injustices occurring, we must look at all angles from which they could possibly be coming from. This is because someone might only be testimonially injust toward a person who is female, others might only act unjustly because of your membership with a historically marginalized race, and so on.  We contend that the better metrics to use for detecting injustices, e.g. testimonial injustice, in medical records are ones which consider intersectionality. Still, we see differences in how these measures show which groups are experiencing privilege, thus we must be careful in understanding the goals of the fairness metrics we use.



\section{Conclusions}
The objective of this empirical study was to investigate the potential benefits of intersectionality in detecting testimonial injustice, using medical records as a real-world application. Demographic parity, differential intersectional fairness, and subgroup fairness were used to examine whether there are differences in the extent of testimonial injustice experienced by individuals based on the intersection of their demographic attributes and if intersectionality helps reveal this. Our results showed (1) when we allow ourselves to use metrics that consider intersectionality, as opposed to sole factors of who a person is, we can better see disparities in how they are treated in terms of detecting testimonial injustice in medical records, (2) there are differences in how someone is treated based on the intersection of their demographic attributes (3) different intersectional fairness metrics do reveal these injustices differently. While demographic parity did not show a clear disparate impact based on gender or race, differential intersectional fairness and subgroup fairness -- two intersectional fairness measures -- revealed that there was disparate treatment based on both gender and race. These findings suggest that intersectionality should be considered when detecting testimonial injustice, especially in medical settings. 

\section{Limitations and Future Work}
\noindent\textbf{Data.} A challenge we faced was that MIMIC-III was unevenly distributed across the races (e.g. ethnicities) for the patients represented. We had significantly more White and Black patients than any other race of people and even still many more White than Black patients. Therefore we continue to express the need for more representative, inclusive, and balanced datasets. Further, the dataset did include ethnic breakdowns, but due to the lack of patients present in those ethnic groups we could not include Caribbean or Middle Eastern patients as well as many other subgroups in our analysis. We would like to use a more comprehensive dataset in the future, potentially from a facility that consistently services marginalized and privileged communities. If we had more time, we would like to partner with a medical facility that regularly serves marginalized and non-marginalized groups, steadily, to develop a dataset which captures more features that could reveal some bias and ensure they are more descriptive (i.e. has\_insurance) to get higher quality data. 

\noindent\textbf{Better Feature Selection and Using More Demographic Features.} To ensure the quality of the aforementioned data, we will perform a causal analysis to identify the specific features that cause testimonial injustice. We anticipate that variables such as age and education level of patients need be included, as these factors have been shown to affect how patients are treated, particularly in the medical field \citep{dunsch2018bias,devoe2009patient}.

\noindent\textbf{Fairness Metrics.} Existing and popular, fairness metrics cannot be generalized to fit in settings where intersectionality must be considered. Another challenge we faced was having a lack of good baselines to use when analyzing intersectional differences. Intersectionality is highly unexplored, in the future we would like to develop our own metric which can be more beneficial in detecting intersectional disparate treatment between individuals.

\noindent\textbf{Additional Analysis.} We plan to conduct additional analysis to understand if specific physicians treat similar patients similarly based on the intersection of their demographic features. Further, we plan to perform statistical significance testing on differences in how patients were treated based on the intersection of their demographic features and the occurrences of specific physicians' use of testimonial unjust terms to other patients.

\section*{Acknowledgements}
This paper is based upon work supported in part by the NSF LSAMP Bridge to the NSF Program on Fairness in AI in Collaboration with Amazon under Award No. IIS-1939743, titled FAI: Addressing the 3D Challenges for Data-Driven Fairness: Deficiency, Dynamics, and Disagreement (Kenya Andrews). This work is also supported in part by the Cisco Research Gift Grant (Lu Cheng). Any opinion, findings, and conclusions or recommendations expressed in this paper are those of the authors and do not necessarily reflect the views of the National Science Foundation, Amazon, or Cisco Research.

\bibliography{anthology,custom}

\pagebreak
\appendix 
\onecolumn
\section{Appendix} \label{sec: append}

\subsection{Normalized Sums of Unjust Terms} \label{ssec: normsumappendixsec}
\begin{figure*}[!htbp]
\centering
\includegraphics[width=0.9\linewidth]{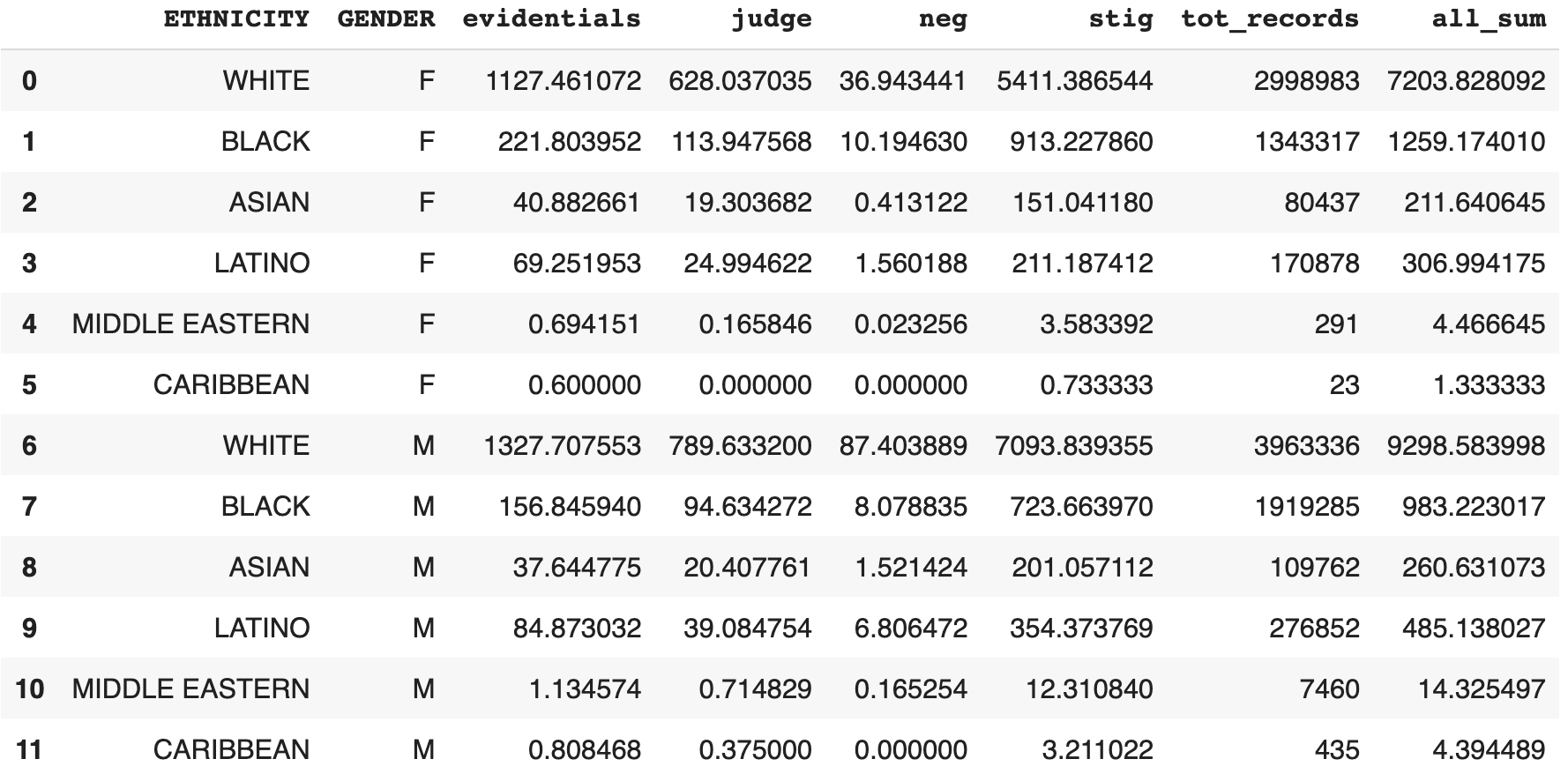}
    \caption{Normalized sums of occurrences of unjust terms for patients based on race and gender. Higher numbers indicate higher counts of terms.}
    \label{fig:interscores}
\end{figure*}

\subsection{Abbreviations}
\begin{table*}[!htbp]
\fontsize{11pt}{12pt}\selectfont
\centering
\begin{tabular}{l|l}
\hline
\textbf{Abbreviation} & \textbf{Full Form} \\ \hline
F & Female\\ \hline
M & Male \\ \hline
W & White \\ \hline
B & Black \\ \hline
A & Asian   \\ \hline
L & Latino   \\ \hline
WF & White Female   \\ \hline
BM & Black Male   \\ \hline
... & ...    
\end{tabular}
\caption{Abbreviations of Demographic Features and their Combinations.} 
\label{tab:legend}
\end{table*}

\pagebreak
\subsection{Intersectional Analysis of Terms}\label{ssec: dist-analysis}
In conducting analysis on the MIMIC-III dataset, we plot the distributions of the occurrences of each term which can lead to testimonial injustice. The position of the peak in the distribution graph provides insight into which subgroups are experiencing a stronger degree of injustice. The more right-skewed the peak of the distribution is, the higher amount of injustice experienced by that particular subgroup. Naturally, the height of the peak speaks to the confidence of the severity to which that subgroup is experiencing injustice based on their word count.

\begin{figure*}[!htbp]
  \centering
\setkeys{Gin}{width=0.5\linewidth}
  \includegraphics{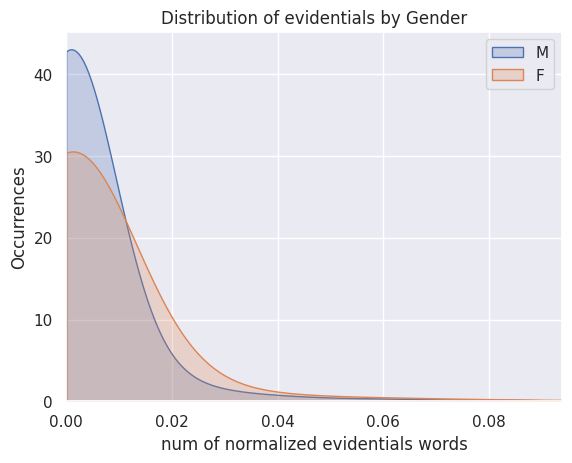}\,%
  \includegraphics{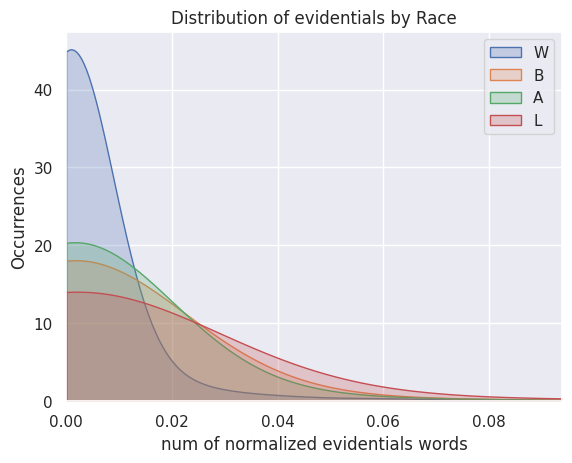}
  \caption{Distribution of Evidential terms in medical notes, refer to legend in Table \ref{tab:legend}  to see the full text of the abbreviated terms. Left: Shows distribution of gender-only. Right: Shows the distribution of the intersection of race-only. }
  \label{fig:evi-graphs-append}
\end{figure*}
\begin{figure*}[!htbp]
\centering
\includegraphics[width=0.5\linewidth]{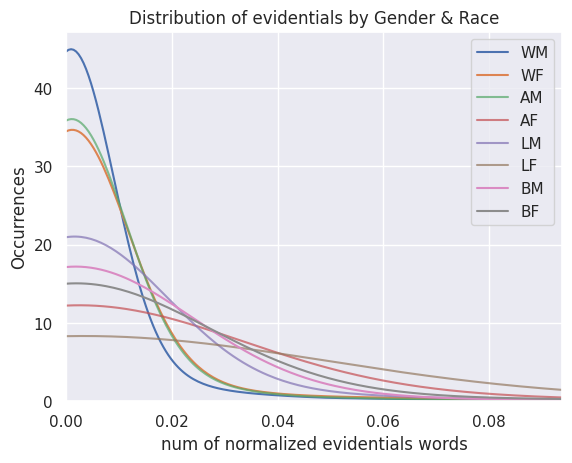}
    \caption{Distribution of Evidential terms considering intersectionality in medical notes, refer to legend in Table \ref{tab:legend} to see the full text of the abbreviated terms. }
    \label{fig:evi-graphs}
\end{figure*}

In comparing Figures \ref{fig:evi-graphs-append} and \ref{fig:evi-graphs} notice in terms of race, Asian patients experience evidential terms the second least, after White patients. Still, Asian Females have the second most highest occurrences of evidential terms, which is a clear contradiction, showing the importance of observing intersectional experiences. In Figure \ref{fig:evi-graphs}, we observe the normalized distribution of evidential terms used for patients across different intersections of races and genders. White men, Asian men, and White females show lower amounts of evidential terms in their records, while Latina females, Asian females, and Black females have higher occurrences of evidential terms in their medical records.  

\begin{figure*}[!htbp]
  \centering
\setkeys{Gin}{width=0.5\linewidth}
  \includegraphics{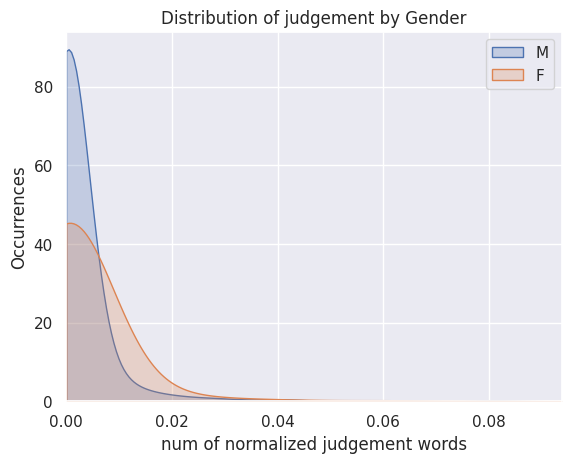}\,%
  \includegraphics{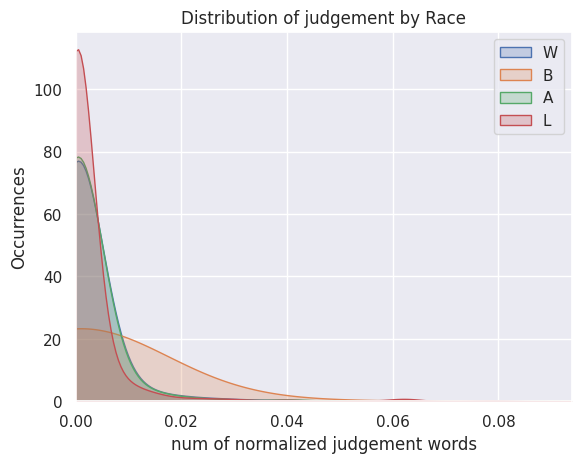}
  \caption{Distribution of Judgement terms in medical notes, refer to legend in Table \ref{tab:legend} to see the full text of the abbreviated terms. Left: Shows distribution of gender-only. Right: Shows the distribution of the intersection of race-only.}
  \label{fig:judge-graphs-append}
\end{figure*}

\begin{figure*}[!htbp]
\centering
\includegraphics[width=0.5\linewidth]{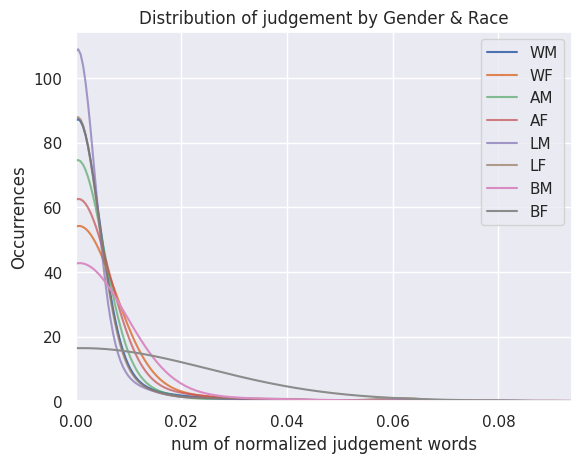}
    \caption{Distribution of Judgement terms in medical notes, refer to legend in Table \ref{tab:legend} to see the full text of the abbreviated terms. Refer to Figure \ref{fig:judge-graphs-append} to see gender-only and race-only graphs.}
    \label{fig:judge-graphs}
\end{figure*}

\pagebreak
From the normalized distributions of the occurrence of judgement terms in the medical records, in Figure \ref{fig:judge-graphs-append} we can observe that female patients as opposed to male patients and Black patients as apposed to the other races, studied here, have the most occurrences of judgement terms. Figure \ref{fig:judge-graphs} emphasizes just how much worse Black women are impacted than any other subgroup. Black men and White women are the next two most vulnerable groups to experiencing judgemental terms in their physicians' notes. Latino men, White men, and Latina females have the least occurrences of judgement terms in their records.

\pagebreak
From Figure \ref{fig:stig-graphs} we observe the distributions of normalized stigmatizing terms used for patients over the intersection of their race and gender. Asian men, followed by Asian females and White males have experienced the least stigmatizing language in the physicians' notes, while Black females and Latino men have been faced with it the most. Figure \ref{fig:evi-graphs-append} suggests that Latino and Black patients receive similar treatment, however, Figure \ref{fig:evi-graphs} highlights that stigmatizing language is more prevalent in the medical records of Black females and Latino males compared to any other subgroups.

\begin{figure*}[!htbp]
  \centering
\setkeys{Gin}{width=0.5\linewidth}
  \includegraphics{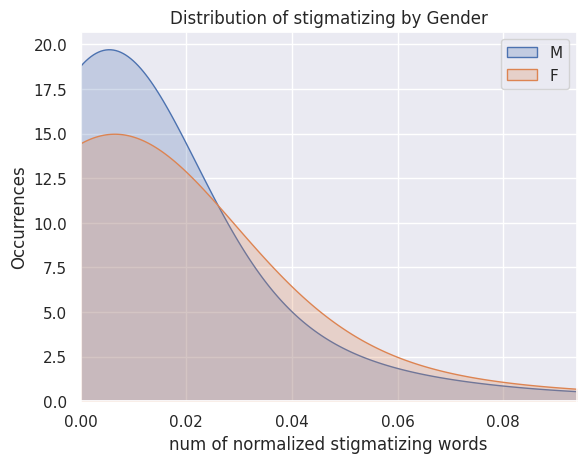}\,%
  \includegraphics{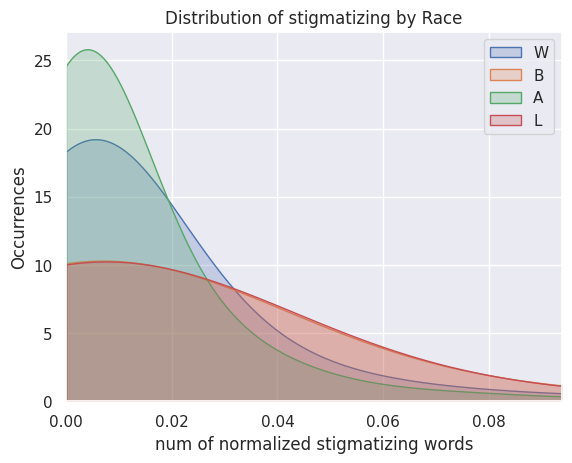}
  \caption{Distribution of Stigmatizing  terms in medical notes, refer to legend in Table \ref{tab:legend} to see the full text of the abbreviated terms. Left: Shows distribution of gender-only. Right: Shows the distribution of the intersection of race-only.}
  \label{fig:stig-graphs-append}
\end{figure*}

\begin{figure*}[!htbp]
\centering
\includegraphics[width=0.5\linewidth]{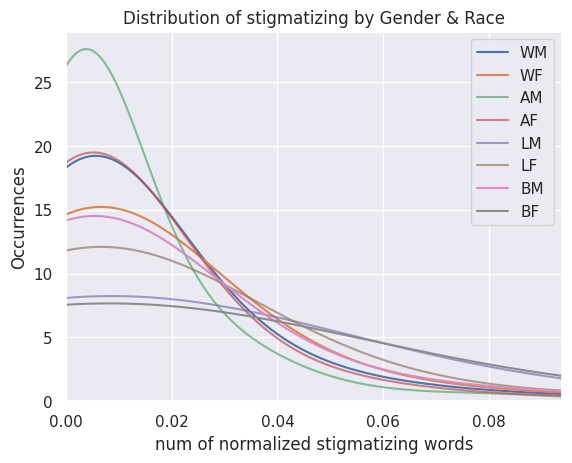}
    \caption{Distribution of Stigmatizing  terms in medical notes, refer to legend in Table \ref{tab:legend} to see the full text of the abbreviated terms. Refer to Figure \ref{fig:stig-graphs-append} to see gender-only and race-only graphs.}
    \label{fig:stig-graphs}
\end{figure*}

To conclude, these graphs specifically their variations show the importance of exploring intersectionality while providing medical care. For example, Black females face challenges that are unique to their intersectional identity as both black and female. This intersectionality can result in compounded experiences of discrimination and marginalization. Furthermore, the fact that Asian and White males consistently occupy the most privileged subgroup highlights systemic inequalities and the need for continued efforts to address these disparities. 

\end{document}